\documentclass[11pt]{article}
\usepackage{amsmath, amsthm, amssymb}    
\usepackage{bridges}			
\usepackage{graphicx}			
\usepackage{times}
\usepackage[colorlinks=true, urlcolor=black, citecolor=black, linkcolor=black]{hyperref}  

\newcommand{\ignorethis } [1] { }

\newcommand{\fignum     } [1] {\ref{#1}}

\newcommand{\fig        } [1] {Figure~\fignum{#1}}

\newcommand{\Reals      }     {{\textrm{I\kern-0.18em R}}}

\newcommand{\change     } [1] {\mbox{{\footnotesize $\Delta$} \kern-3pt}#1}

\newcommand{\pctab}{\hspace{0.2in}}

\graphicspath{{figures/}}

\title{Heesch Numbers of Unmarked Polyforms}

\author{Craig S. Kaplan \\
School of Computer Science, University of Waterloo, Ontario, Canada; csk@uwaterloo.ca} 

\date{}					

\begin{document}

\maketitle

\begin{abstract}

A shape's Heesch number is the number of layers of copies of the
shape that can be placed around it without gaps or overlaps.
Experimentation and exhaustive searching have turned up examples of
shapes with finite Heesch numbers up to six, but nothing higher.
The computational problem of classifying simple families of shapes
by Heesch number can provide more experimental data to fuel our
understanding of this topic.  I present a technique for computing 
Heesch numbers of non-tiling polyforms using a SAT solver, and the
results of exhaustive computation of Heesch numbers up to 
19-ominoes, 17-hexes, and 24-iamonds.

\end{abstract}

\section{Introduction}

Tiling theory is the branch of mathematics concerned with the
properties of shapes that can cover the plane with no gaps or
overlaps.  It is a topic rich with deep results and open problems.
Of course, tiling theory must occasionally venture into the study
of shapes that \textit{do not} tile the plane, so that we might
understand those that do more completely.

If a shape tiles the plane, then it must be possible to surround
the shape by congruent copies of itself, leaving no part of its
boundary exposed.  A circle clearly cannot tile the plane, because
neighbouring circles can cover at most a finite number of points
on its boundary.  A regular pentagon also cannot be surrounded by
copies of itself: its vertices will always remain exposed.

\begin{figure}
\begin{center}
\includegraphics[width=5in]{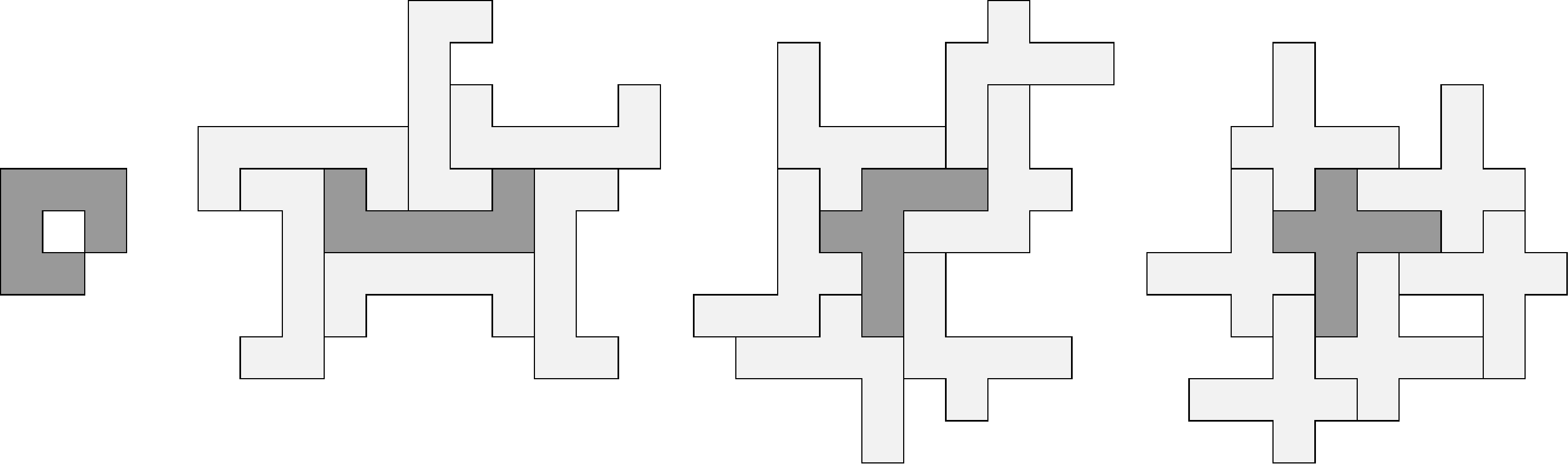}
\end{center}
\caption{\label{fig:7}The four non-tiling heptominoes.
	The shape on the left has a hole and cannot be surrounded.
	The other three can be fully surrounded by copies, but in the
	rightmost shape the copies will necessarily enclose a hole.}
\end{figure}

However, the converse is not true:
there exist shapes that can be fully surrounded by copies of themselves,
but for which no such surround can be extended to a tiling.  For example,
there are 108 heptominoes (shapes formed by gluing together seven squares),
of which four, shown in \fig{fig:7}, are known
not to tile the plane. One of them contains an internal hole and can be 
discarded immediately.  As it happens, the other three can all be
surrounded.  In the
middle two cases, the shape and its surrounding copies are simply
connected. On the right, the surrounding tiles leave behind an internal
hole, and no alternative surround can eliminate that hole.  

\begin{figure}
\begin{center}
\includegraphics[width=3in]{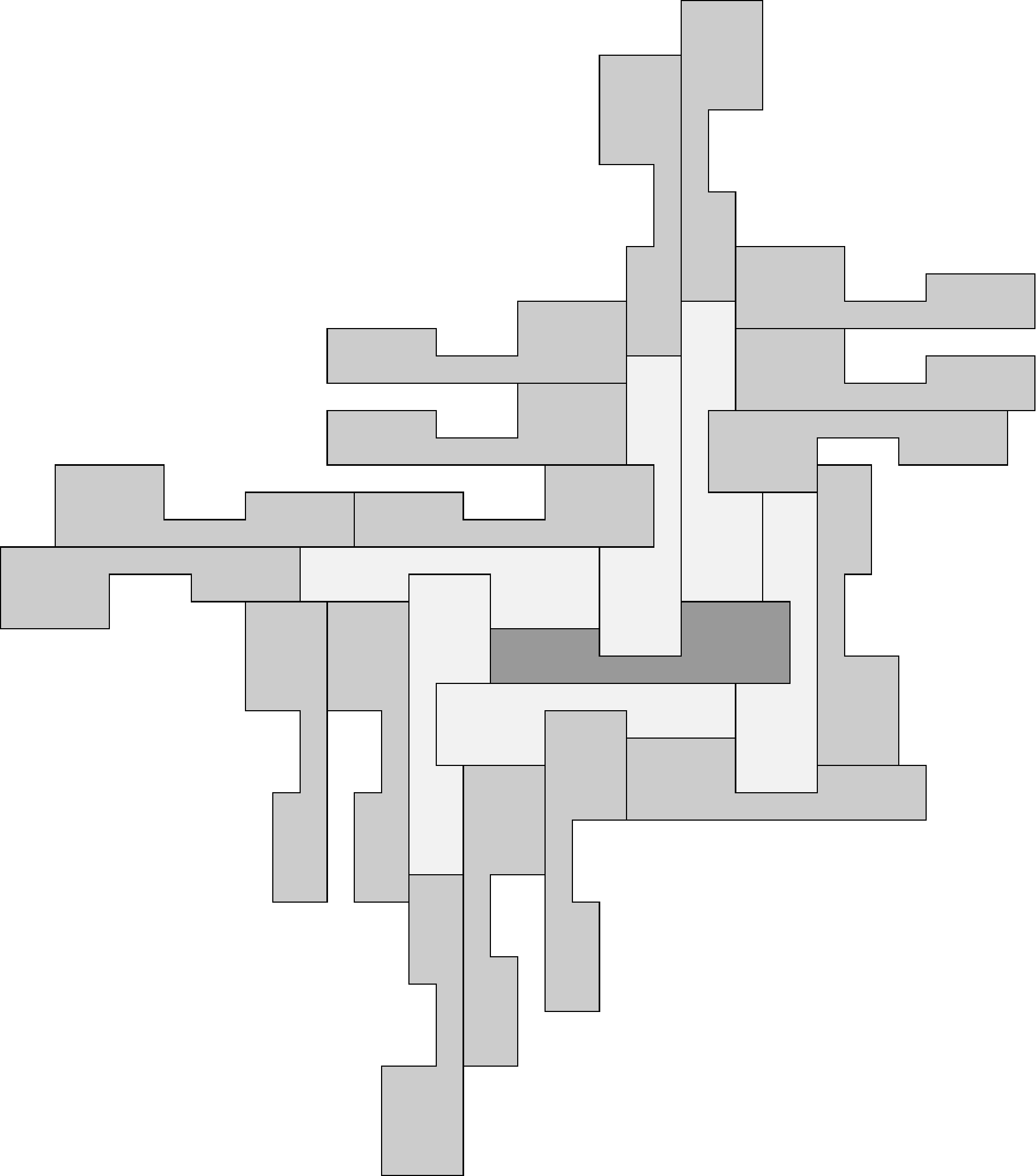}
\end{center}
\caption{\label{fig:fontaine}A 23-omino that can be surrounded by
	two layers of copies of itself, but not more.}
\end{figure}

There is no a priori reason why a given non-tiling shape might not
be surroundable by two, three, or more layers of copies of itself.
The illustrations in \fig{fig:7} provide lower bounds for
the numbers of layers for these shapes; that they also represent upper
bounds must
be proven by enumerating all possible surrounds, and showing that
none of them may be further surrounded.  Other shapes might permit
more layers.  For example, the 23-omino shown  in \fig{fig:fontaine},
due to Fontaine~\cite{Fontaine}, can be surrounded by two layers
but not more.  How far can this process be extended?

A shape's \textit{Heesch number} is the number of times it can be
surrounded with complete layers of congruent copies of itself (I
will offer a precise definition in the next section).  If the shape
tiles the plane, its Heesch number is defined to be infinity.
\textit{Heesch's problem} asks which positive integers are Heesch
numbers; that is, for which $n>0$ does there exist a shape with
Heesch number $n$?  


Very little is known about the solution to Heesch's
problem.  Writing in 1987, Gr\"unbaum and Shephard were not aware
of any examples with finite Heesch number greater than 1~\cite[Section
3.8]{GS}.  After that, a few isolated examples were found with
Heesch numbers up to 4~\cite{Mann2004}.  Mann and Thomas performed
a systematic computer search of marked polyforms
(polyominoes, polyhexes, and polyiamonds, with edges decorated with
geometric matching conditions), yielding new
examples and pushing the record to 5~\cite{MT2016}.  In 2021
Ba{\v{s}}i{\'c} finally broke this record, demonstrating a figure
with Heesch number 6~\cite{Basic2021}.

The study of Heesch numbers can shed light on some of the deepest problems
in tiling theory.  In particular, the \textit{tiling problem} asks, for
a given set of shapes, whether they admit at least one tiling of the plane.
The tiling problem is known to be undecidable for general sets of 
shapes~\cite{Berger}, but its status is open for a set consisting of a
single shape $S$.  If there were an upper bound $N$ on finite Heesch numbers,
then the tiling problem would be decidable, at least when there are only
finitely many ways that two copies of $S$ may be adjacent~\cite{Chaim}.
The algorithm would involve trying the finitely many ways of surrounding $S$ 
with $N+1$ layers of copies of itself.  If you succeed, then you have 
exceeded the maximum finite Heesch number and $S$ must tile the plane.
If you fail, then $S$ evidently does not tile.  To that end,
more experimental data revealing which Heesch numbers are possible, even 
for limited classes of shapes, could be useful in understanding whether
such an upper bound might exist.

In this article I report on a complete enumeration of Heesch numbers
of unmarked polyforms, up to 19-ominoes, 17-hexes, and 24-iamonds.
This enumeration comprises approximately 4.16 billion non-tilers, extracted
from enumerations of all free polyforms of those sizes.  
Respecting a slight difference of opinion among researchers, I compute
two variations of Heesch numbers: one where tiles may form holes in the
outermost layer, and one where a shape and all its surrounding layers must
be simply connected.
This enumeration does not shatter the existing records for Heesch numbers,
but it does provide a store of new examples of shapes with non-trivial
Heesch numbers.  Some, like a 9-omino with Heesch number 2 
(\fig{fig:ominoes}) and a 
7-hex with Heesch number 3
(\fig{fig:hexes}), are interesting because of the complex behaviour
exhibited by relatively simple shapes.  The enumeration also uncovered 
seven new examples with Heesch number 4.

Apart from the tabulation and specific examples, the other main contribution
of this work lies in the use of a SAT solver to compute Heesch numbers.
Because polyominoes, polyhexes, and polyiamonds are subsets of ambient
regular tilings of the plane, it is possible to reduce the geometric 
problem of surroundability to the logical problem of satisfiability
of Boolean formulas.  A SAT solver can optimize its
search of the exponential space of possible solutions, avoiding the risk
of ``backtracking hell''.  This formulation leads to a very reliable
algorithm, whose performance degrades only on the rare shapes that
actually have high Heesch numbers.

\section{Mathematical background}

Although Heesch's problem grew out of tiling theory, most of the
language, techniques, and results of tiling theory are not needed
within the scope of this article and will be omitted.  Readers
interested in the topic should consult Gr\"unbaum and Shephard's
book~\cite{GS}, which remains the standard reference.  In this section
I will formalize the definition of a shape's Heesch number and review
marked and unmarked polyforms.

\subsection{Heesch numbers}
\label{sect:heesch}

Let $C$ and $S$ be simple shapes in the plane, i.e., 
topological discs.  We say that $C$ can be \textit{surrounded by}
$S$ if there exists a set of shapes $\{S_1,\ldots,S_n\}$ with the 
following properties:

\begin{enumerate}
\item Each $S_i$ is congruent to $S$ via a rigid motion in the plane;
\item The shapes in the set 
	$\{C,S_1,\ldots,S_n\}$ have pairwise disjoint interiors;
\item The boundary of each $S_i$ shares at least one point with the
	boundary of $C$.
\item The boundary of $C$ lies entirely within the interior of 
	the union of $C$ and all the $S_i$.
\end{enumerate}

The second condition forces the shapes not to overlap, except on their
boundaries.  The third condition forces every $S_i$ to be useful in 
covering the boundary of $C$.  The fourth condition ensures that $C$
is completely surrounded.

If, furthermore, the union of $C$ and the $S_i$ is simply connected,
we say that $C$ can be surrounded by $S$ \textit{without holes}.  In
tiling theory, a finite union of non-overlapping shapes whose union is a 
topological disc is also known as a \textit{patch}, a term I will use
here.  On the other hand, I will use the more general term \textit{packing}
when shapes are known to be non-overlapping but when their union may or
may not contain holes.

We formalize the notion of layers
by defining the \textit{coronas} of $S$.
We define the \textit{$0$-corona} of $S$ to be the singleton
set $\{S\}$.
Setting $C=S$ above, if $S$ can
be surrounded by itself then the tiles that make up that surround are one 
possible \textit{$1$-corona} of $S$.  In general, if we have a nested
sequence of $k$-coronas for $k=0,\ldots,n-1$, all without holes, and the 
patch created from the union of all
of these coronas can itself be surrounded by $S$, then the copies of $S$
making up the surround constitute an \textit{$n$-corona}.

The \textit{Heesch number} of a shape $S$ is the largest $n$ for
which $S$ has an $n$-corona.  If $S$ tiles the plane,
then by definition it is possible to build an $n$-corona for every
positive integer $n$, and we define its Heesch number to be infinity.
If we wish to be concise, we will simply say that $S$ \textit{has $H=n$}.

The definitions above require that for a shape to have Heesch number
$n$, each $k$-corona for $k=1,\ldots,n-1$ surround its predecessor
without holes. But it leaves the status of the outermost corona
ambiguous.  Most researchers require that a shape's $n$-corona be
hole-free in order to regard the shape as having $H=n$, but some
permit the $n$-corona to have holes.  In this article I will 
remain neutral on this point, and report separate results with
and without holes in the outer corona. 
To that end, I will say that a
shape has $H_c=n_1$ and $H_h=n_2$ to distinguish its Heesch numbers
when holes are or are not permitted in the outer corona, respectively.
In any case, we must always have either $H_c=H_h$ or $H_c=H_h-1$,
so this difference of opinion cannot affect results too dramatically.
(Note that permitting a hole in the outermost corona raises the alarming 
possibility that the hole could be filled with additional tiles, forcing
us to consider the validity of a subsequent corona made from multiple
disjoint pieces!)

\subsection{Polyforms}

A \textit{polyform} is a shape constructed by
gluing together multiple copies of some simple polygonal building
block along their edges.  Usually we require that the assembly be
\textit{edge-to-edge}: no vertex of one copy of the building block may lie
in the interior of the edge of another copy.  The most famous
polyforms are the \textit{polyominoes}, constructed from glued-together
squares.  We speak more specifically of \textit{$n$-ominoes} as unions of
$n$ squares, so that, for example, the 4-ominoes (or tetrominoes) are the
familiar Tetris pieces.  In this article I will also consider 
\textit{polyhexes} and \textit{polyiamonds}, formed from unions of
regular hexagons and unions of equilateral triangles, respectively, 
and refer more precisely to
\textit{$n$-hexes} and \textit{$n$-iamonds} as needed.

Simple polyforms are an attractive domain in which to compute Heesch
numbers.  They can be explored exhaustively by enumerating the
finite number of distinct $n$-forms for each successive $n$.  The
edge-to-edge constraint often reduces a continuous geometric problem
to a combinatorial one, and in the technique presented here, even
the combinatorial structure will be distilled into a problem in
Boolean satisfiability.  Still, polyforms can expose many of the
core behaviours of shapes more generally.  Conceivably one could
establish an upper bound on Heesch numbers of, say, 
polyominoes, while leaving Heesch's problem open more generally;
but in the meantime, these calculations can yield a trove of interesting
data.

In a \textit{marked} polyform, the edges of a polyform are assigned
symbolic labels, and a binary relation over labels determines which
pairs of edges may be placed side-by-side in neighbouring copies
of the polyform.  A simple system of labels involves marking some edges
with a ``bump'', some with a corresponding ``nick'', and leaving all
others flat.  Flat edges can only meet other flat edges,
and bumps must be adjacent to nicks.  Mann and Thomas computed
Heesch numbers of simple polyforms with markings of this form~\cite{MT2016}.
They began with a
small family of low-order polyominoes, polyhexes, and polyiamonds, 
enumerated all possible assignments of bumps and nicks to their edges, and
computed the Heesch numbers of the resulting shapes using a recursive
search with backtracking.  Their search yielded a number of examples with
Heesch numbers up to 5.  However, the majority of their efforts produced
inconclusive results: they either failed to produce a finite Heesch
number in the time allotted to each shape, or terminated the computation
at five coronas.
The main reason for this
deficiency is that they did not have an effective procedure for first
computing whether a marked polyform tiles the plane.  Most of their
inconclusive results are likely to be shapes with Heesch numbers that
are infinite, rather than high-but-finite.

To my knowledge, no previous work has sought to compute Heesch
numbers of \textit{unmarked} polyforms.  Myers tabulated information
about polyominoes, polyhexes, and polyiamonds that tile the
plane~\cite{Myers}.  He determined whether polyforms tiled in
progressively more intricate ways, measuring the
\textit{isohedral number} of tilers (roughly speaking, the number
of copies of the tile that must be glued together to produce a patch
that tiles in a relatively simple way).  Each of his tables includes
a single column labelled ``non-tilers''.  This article
sorts that columns into multiple bins organized by
Heesch number, effectively tabulating the progressively more intricate
ways in which polyforms \textit{do not} tile.  Myers's software is
remarkably efficient, requiring on average a fraction of a millisecond
on modern hardware to classify a given polyform.  I use his software
to produce initial lists of non-tilers
for Heesch number computation, thereby avoiding the needless
construction of coronas for shapes that have infinitely many of them.

\section{Computing Heesch numbers with a SAT solver}

In this section I show how to reduce the problem of computing a polyform's
Heesch number to evaluating the satisfiability of a sequence of Boolean 
formulas.  At a high level, each 
formula encodes whether a given polyform has Heesch number at least $n$
(with slight variations depending on whether to allow holes).  I check
the satisfiability of these formulas for increasing values of $n$ until
I find one that is unsatisfiable, indicating the non-existence of a
corona of a given level.

Every formula will be expressed in \textit{conjunctive normal form} (CNF)
as a conjunction of clauses, each of which is a disjunction of variables
or their negations.  That is, each clause ORs together any number of 
Boolean variables or their negations, and the entire formula is an AND
of clauses.  I use the standard operators $\vee$ for OR, $\wedge$ for
AND, and $\neg$ for NOT.  I will also allow clauses to be written 
using an implication operator with a single variable on the left, 
converting $P\rightarrow Q$ to $\neg P\vee Q$ as needed.

To simplify the exposition, I will limit the development here exclusively 
to polyominoes.  In the next section I will describe the modifications
that are necessary to support polyhexes and polyiamonds.

\subsection{Developing the base formula}

Because our shapes will always meet edge-to-edge, we can assume that
they will occupy cells in a conceptually infinite grid of squares,
indexed by $(x,y)$ pairs of integer coordinates.  For a given cell
$p=(x,y)$, we define $N_8(p)$, the 8-neighbourhood of $p$, to be the
set of cells horizontally, vertically, or diagonally adjacent to $p$.
Now let $S$ be an
$m$-omino whose Heesch number we wish to compute.  We describe $S$
as a set of cells $\{(x_1,y_1),\ldots,(x_m,y_m)\}$, translated so that
$(0,0)\in S$.  We will also make use of the \textit{halo} of $S$, written
$\textsc{Halo}(S)$, the set of grid cells $p$ for which 
$p \notin S$ but $N_8(p)\cap S\neq \varnothing$.  That is, 
$\textsc{Halo}(S)$ consists of a ring of cells around the boundary of $S$.

Ignoring symmetry, a polyomino has eight distinct rotated and
reflected orientations, which can be represented by $2\times 2$
matrices with entries in $\{-1,0,1\}$.  We must also track translations
of polyominoes by integer vectors $(\Delta x, \Delta y)$.  Any
possible transformed copy of $S$ can therefore be identified with six
(usually small) integers that define an affine transformation $T$.
Two transformed shapes  $T_1(S)$ and $T_2(S)$ are \textit{adjacent} 
if they occupy neighbouring cells but do not overlap; 
that is, $T_1(S)\cap T_2(S)=\varnothing$, but
$T_1(S)\cap \textsc{Halo}(T_2(S)) \neq \varnothing$.  For a fixed $S$,
I will also refer to $T_1$ and $T_2$ as adjacent in this context.

We are particularly interested in finite sets of transformations
$\mathcal{T}_k$,
containing every possible $T$ for which $T(S)$ might 
be part of a $k$-corona of $S$.  We can define these sets recursively
by setting $\mathcal{T}_0$ to be a singleton set containing the identity
transformation,
and each subsequent $\mathcal{T}_k$ to be every transformation $T$ adjacent
to some $T'\in\mathcal{T}_{k-1}$.  Every $k$-corona of $S$, if one exists,
must consist of copies of $S$ transformed by a subset of $\mathcal{T}_k$.

We are now ready to define two classes of Boolean variables: 
cell variables and shape variables. 
For every $p=(x,y)$ in the grid, the \textit{cell variable}
$c_p$ is true if and only if $p$ is covered by a transformed copy of $S$.
For every affine transformation $T$ and every integer $k\ge 0$, the 
\textit{shape variable} $s_{T,k}$ is true if and only if the transformed
shape $T(S)$ is used as part of the $k$-corona in a packing of copies of
$S$.

Given an integer $n>0$, we can at last write down a Boolean formula $F_n$
whose satisfiability implies that $S$ has an $n$-corona.
$F_n$ is the conjunction of a large number of clauses, belonging to seven
distinct classes.  The clauses are listed in full in \fig{fig:clauses},
along with intuitive explanations of their meanings.
Informally, we see that the 0-corona is activated by fiat, which
in turn demands that its halo cells all be occupied by adjacent
shapes.  Additional clauses force those adjacent shapes to belong
to the 1-corona, and to be pairwise disjoint.  A similar process
plays out in each subsequent corona before the last one: shapes in
the corona tag their halo cells, thereby recruiting new neighbours to
surround them.  The shapes in the outermost corona are left 
partially exposed to empty space.

\begin{figure}
\begin{center}
\begin{tabular}{l|l}
Clause and quantifiers & Explanation \\
\hline
$s_{I,0}$  &
\begin{minipage}[t]{3in}
The 0-corona is always used.
\end{minipage} \\
\\
\begin{minipage}[t]{2in}
	$s_{T,k}\rightarrow c_p$ \\
	\hspace*{0.25in} For all $0\le k \le n$ \\
	\hspace*{0.25in} For all $T\in \mathcal{T}_k$ \\
	\hspace*{0.25in} For all $p\in T(S)$
\end{minipage} &
\begin{minipage}[t]{3in}
If a copy of $S$ is used, then its cells are used. 
\end{minipage} \\
\\
\begin{minipage}[t]{2in}
	$c_p \rightarrow s_{T_1,k_1} \vee \ldots \vee s_{T_m,k_m}$ \\
	\hspace*{0.25in} For all $0\le k_i \le n$ \\
	\hspace*{0.25in} For all $T_i\in \mathcal{T}_{k_i}$ \\
	\hspace*{0.25in} Where $p\in T_i(S)$
\end{minipage} &
\begin{minipage}[t]{3in}
If a cell is used, then some copy of $S$ must use it. 
\end{minipage} \\

\\
\begin{minipage}[t]{2in}
	$s_{T,k} \rightarrow c_q$ \\
	\hspace*{0.25in} For all $0\le k \le n-1$ \\
	\hspace*{0.25in} For all $T\in \mathcal{T}_k$ \\
	\hspace*{0.25in} For all $q\in \textsc{Halo}(T(S))$
\end{minipage} &
\begin{minipage}[t]{3in}
If a copy of $S$ is used in an interior corona (a $k$-corona for 
$k<n$), then that copy's halo cells must be used. 
\end{minipage} \\
	
\\
\begin{minipage}[t]{2in}
	$s_{T_1,k_1} \rightarrow \neg s_{T_2,k_2}$ \\
	\hspace*{0.25in} For all $0\le k_1, k_2 \le n$ \\
	\hspace*{0.25in} For all $T_1\in \mathcal{T}_{k_1}$ and $T_2\in\mathcal{T}_{k_2}$ \\
	\hspace*{0.25in} Where $(T_1,k_1)\neq(T_2,k_2)$ \\
	\hspace*{0.25in} And $T_1(S)\cap T_2(S)\neq \varnothing$
\end{minipage} &
\begin{minipage}[t]{3in}
Used copies of $S$ cannot overlap.
\end{minipage} \\

\\
\begin{minipage}[t]{2in}
	$s_{T,k} \rightarrow s_{T_1,k-1}\vee \ldots \vee s_{T_m,k-1}$ \\
	\hspace*{0.25in} For all $1\le k \le n$ \\
	\hspace*{0.25in} For all $T_i\in \mathcal{T}_{k-1}$ \\
	\hspace*{0.25in} Where $T_i$ is adjacent to $T$
\end{minipage} &
\begin{minipage}[t]{3in}
If a copy of $S$ is used in a $k$-corona, it must be adjacent to a copy
in a $(k-1)$-corona
\end{minipage} \\

\\
\begin{minipage}[t]{2in}
	$s_{T_1,k} \rightarrow \neg s_{T_2,m}$ \\
	\hspace*{0.25in} For all $2\le k \le n$ \\
	\hspace*{0.25in} Where $T_1\in \mathcal{T}_k$ \\
	\hspace*{0.25in} For all $0 \le m \le k-2$ \\
	\hspace*{0.25in} For all $T_2\in \mathcal{T}_m$ \\
	\hspace*{0.25in} Where $T_2$ is adjacent to $T_1$
\end{minipage} &
\begin{minipage}[t]{3in}
If a copy of $S$ is used in a $k$-corona, it cannot be adjacent to a copy
in an $m$-corona for $m<k-1$.
\end{minipage} \\

\end{tabular}
\end{center}
\caption{\label{fig:clauses}The clauses that make up the Boolean formula
	$F_n$, which is satisfiable if a shape $S$ has an $n$-corona.}
\end{figure}

The formula $F_n$ can be given to a SAT solver, a program that consumes a
Boolean formula and determines whether any assignment of true or false to
its variables makes the entire formula true.  If the solver
reports that
$F_n$ is satisfiable, then the coronas of $S$ can be read directly from
the true variables $s_{T,k}$ in the satisfying assignment. I iteratively
construct and 
check $F_n$ for each $n\ge 1$ in turn; an unsatisfiable $F_n$ implies
that $S$ has Heesch number $n-1$.  Unfortunately, $F_n$ does not contain a
strict superset of the clauses of $F_{n-1}$, and must be constructed
starting from scratch.

\subsection{Suppressing holes}

If $F_n$ is satisfiable, then the subset of shapes out to the
$(n-1)$-corona will be a simply connected patch: every shape's halo must
be filled, and so no pockets of empty space can be left behind.
However, there is nothing to prohibit holes from forming between
shapes in the $n$-corona.  Thus the algorithm above can compute
only whether $S$ has $H_h=n$.  If we wish to compute the hole-free
Heesch number $H_c$, then we must suppress all holes in the outermost
corona.

\begin{figure}
\begin{center}
\includegraphics[width=4in]{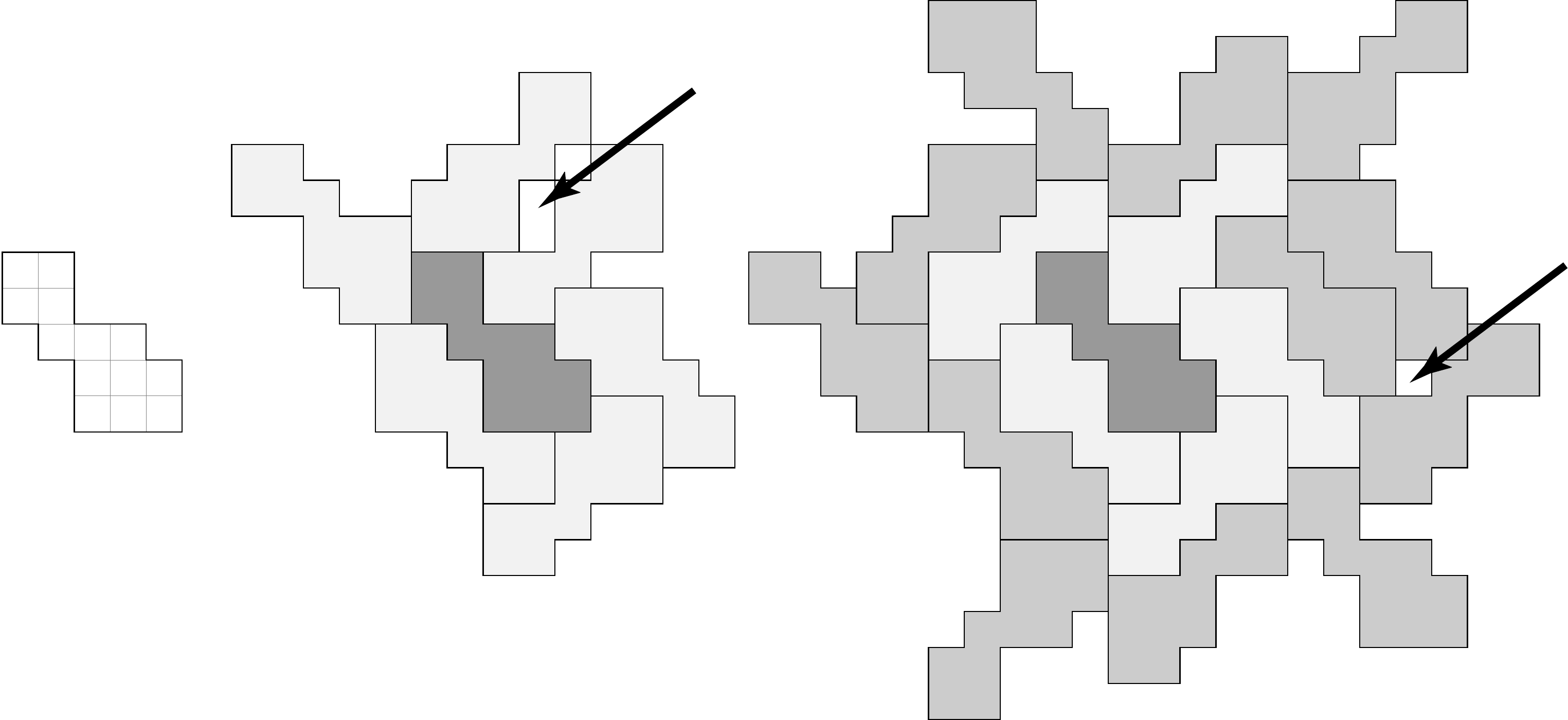}
\end{center}
\caption{\label{fig:holes}A non-tiling 13-omino (left) that
	demonstrates the problem of detecting holes in the outermost corona.
	The middle illustration shows a 1-corona where two adjacent shapes
	enclose holes (one is indicated by an arrow).  These holes can be
	suppressed by including a clause forbidding the two shapes from 
	both being used.  On the right, the 2-corona includes a hole
	bounded by three copies of the shape.
	Such holes are difficult to prevent, and
	are explicitly forbidden after the fact if they are found.}
\end{figure}

Most such holes that might arise are relatively simple, and can be
suppressed easily.  These are holes that are completely enclosed
by a pair of adjacent shapes in the $n$-corona
(\fig{fig:holes}, centre).  I precompute all
pairs of transforms $T_1, T_2\in\mathcal{T}_n$ for which $T_1$ is adjacent
to $T_2$ but $T_1\cup T_2$ is not simply
connected.  When constructing $F_n$, I treat such adjacencies as
illegal, and add clauses of the form $s_{T_1,n}\rightarrow\neg s_{T_2,n}$
to prevent them.

However, it is also possible for the $n$-corona to contain a hole
enclosed by three or more different copies of $S$ (\fig{fig:holes},
right).  It would be prohibitive to precompute and suppress all
possible holes formed by subsets of $\mathcal{T}_n$.  Fortunately,
such holes are exceedingly rare and can be eliminated one at a time
as they arise, using a standard trick from discrete optimization.
If I am trying to compute a shape's hole-free Heesch number, and
$F_n$ is reported as satisfiable, I ``draw'' the implied packing by
assigning symbolic colours to the grid cells in a 2D image, with colours
that index the transformed copies of $S$.  A simple algorithm such 
as flood filling can 
then search the packing for holes.  If none are found, then $S$ has
$H_c\ge n$ and the algorithm proceeds to testing $F_{n+1}$.  If a
hole is found, its boundary will be made up of cells belonging to
shapes transformed by some set $\{T_1,\ldots,T_m\}\subset \mathcal{T}_n$.
I add a clause $\neg s_{T_1,n} \vee \ldots \vee \neg s_{T_m,n}$,
designed to prevent this precise hole, and re-run the SAT solver.
By repeating his process, eventually we 
will either find a hole-free solution, or the solver
will report the enriched $F_n$ as unsatisfiable, implying that $S$
has $H_c<n$.  Unfortunately, verifying that a patch is simply
connected is necessary and potentially expensive; after initial
preprocessing, it is the only part of the process that relies on
the actual geometry of the problem rather than its reduction to
Boolean logic.  I am not aware of an effective way to design $F_n$ to force
the $n$-corona to be simply connected at the outset.

\section{General polyforms}

The geometry of polyominoes makes them easy to work with computationally,
and simplifies the development of the previous section.  All the
geometric computations above can be represented quite compactly in
software.  If we assume that we will not enumerate beyond 23-ominoes
(already an ambitious goal!), and that Heesch numbers will not
exceed 5, then any conceivable set of coronas will fit inside a
$256\times 256$ grid, meaning that a cell coordinate can fit in a
single signed byte.  By the same token, a transformation can easily fit
in 32 bits: at a minimum, we require eight bits each for the coordinates of
the translation, and three more to select a
combination of rotation and reflection.  Furthermore, any copy of
a shape $S$ can be represented implicitly via its transformation, meaning
that construction of $F_n$ can be carried out entirely with 32-bit
integers, regardless of the size of $S$.  It is only when checking whether
a patch is simply connected that I resort to instantiating a large
grid and drawing copies of $S$ in it.

The SAT reduction above can be adapted to other classes of polyforms,
provided that they are expressible as subsets of a fixed ambient
tiling.  That easily encompasses the regular tilings by hexagons and
equilateral triangles, giving us polyhexes and polyiamonds.  It
rules out, for example, shapes formed from edge-to-edge
assemblies of isosceles right triangles, sometimes known as
\textit{polyabolos} or \textit{polytans}.  Of course, even with 
polyhexes and polyiamonds we would like to keep the representation of
shapes and transformations simple, compact, and discrete.  The solution
is to express all coordinates relative to non-standard basis vectors.
This trick is fairly common when working with hexagonal grids in
software, but I will summarize the approach here.

\subsection{Polyhexes}

\begin{figure}
\begin{center}
\includegraphics[width=4in]{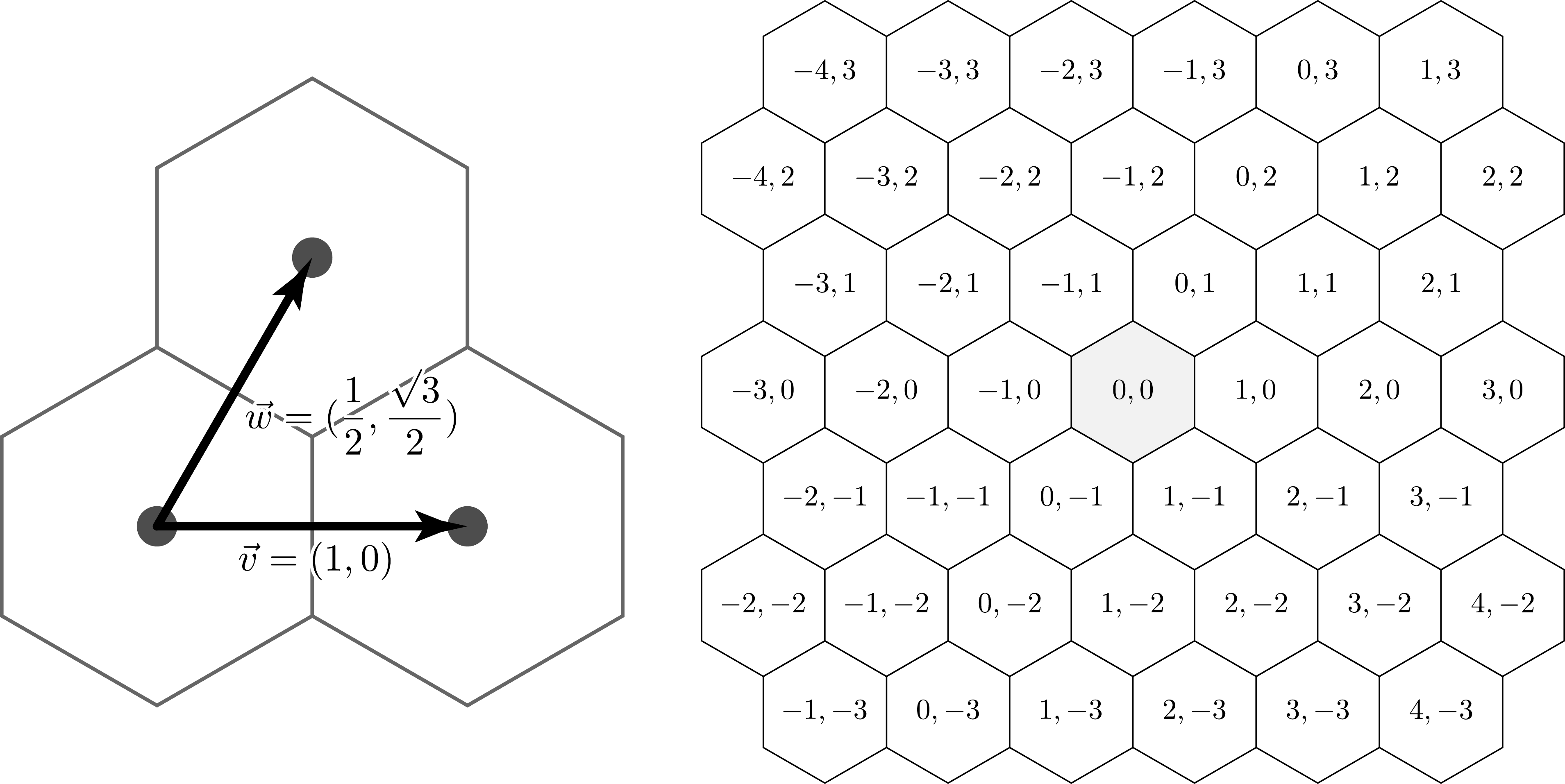}
\end{center}
\caption{\label{fig:hexgrid}The basis on the left allows every cell in
	an infinite hexagonal tiling to be assigned a unique pair of 
	integer coordinates.}
\end{figure}

The cells in a hexagonal grid can be assigned integer coordinates
in a basis with vectors $\vec{v}=(1,0)$ and 
$\vec{w}=(\frac{1}{2},\frac{\sqrt{3}}{2})$,
connecting a hexagon centre to the centres of two of its neighbours.  The
basis is illustrated in \fig{fig:hexgrid}, together with a portion
of a grid labelled with coordinate pairs.  

A hexomino has a maximum of 12 distinct orientations, six direct and six
reflected.  They are generated by a transformation $A$ that rotates by
$60^\circ$ about the origin, and a transformation $B$ that reflects across
the $\vec{v}$ axis.  Working in the basis $\{\vec{u},\vec{v}\}$, these
transformations have simple representations as matrices with integer
entries:

\[ 
	A = \begin{bmatrix} 0 & -1 \\ 1 & 1 \end{bmatrix}, \hspace*{0.1in}
	B = \begin{bmatrix} 1 & 1 \\ 0 & -1 \end{bmatrix}
\]

The products $A^iB^j$ for $i=0,\ldots ,5$ and $j=0,1$ yield matrices
for all 12 orientations which, like their square counterparts, all
have entries in $\{-1,0,1\}$.  These can be combined with translations
by vectors with integer coordinates to represent all possible
transformations of a polyhex.

To construct the halo of a polyhex $S$, we must consider every cell
in the \textit{6-neighbourhood} (and not the 8-neighbourhood) of a
given cell.  These six neighbours can easily be found by offsetting
the coordinates of a cell by the six coordinate pairs in the ring
around $(0,0)$ in \fig{fig:hexgrid}.  The revised definition of 
$\textsc{Halo}(S)$ also affects the definition of adjacency, and
by extension  a number of the clauses that make up $F_n$.

When suppressing holes, verifying that a packing of polyhexes is simply
connected also depends on the distinct topology of the hexagonal
grid.  It is still possible to draw the packing directly into a
square image using the cells' integer coordinates and to use a flood
fill to detect holes.  But unlike the square case, 
after filling an empty grid cell the algorithm
must walk recursively to the empty cells in its 6-neighbourhood.

\subsection{Polyiamonds}

Polyiamonds are slightly more complicated than polyominoes or polyhexes,
in that there are two possible orientations for cells in the infinite
tiling by equilateral triangles.
So, for example, translations cannot simply bring any
triangle into correspondence with any other---they must respect orientation.
I build a somewhat exotic sparse integer representation of the triangular
grid that harnesses the hexagonal representation described above.

\begin{figure}
\begin{center}
\includegraphics[width=4in]{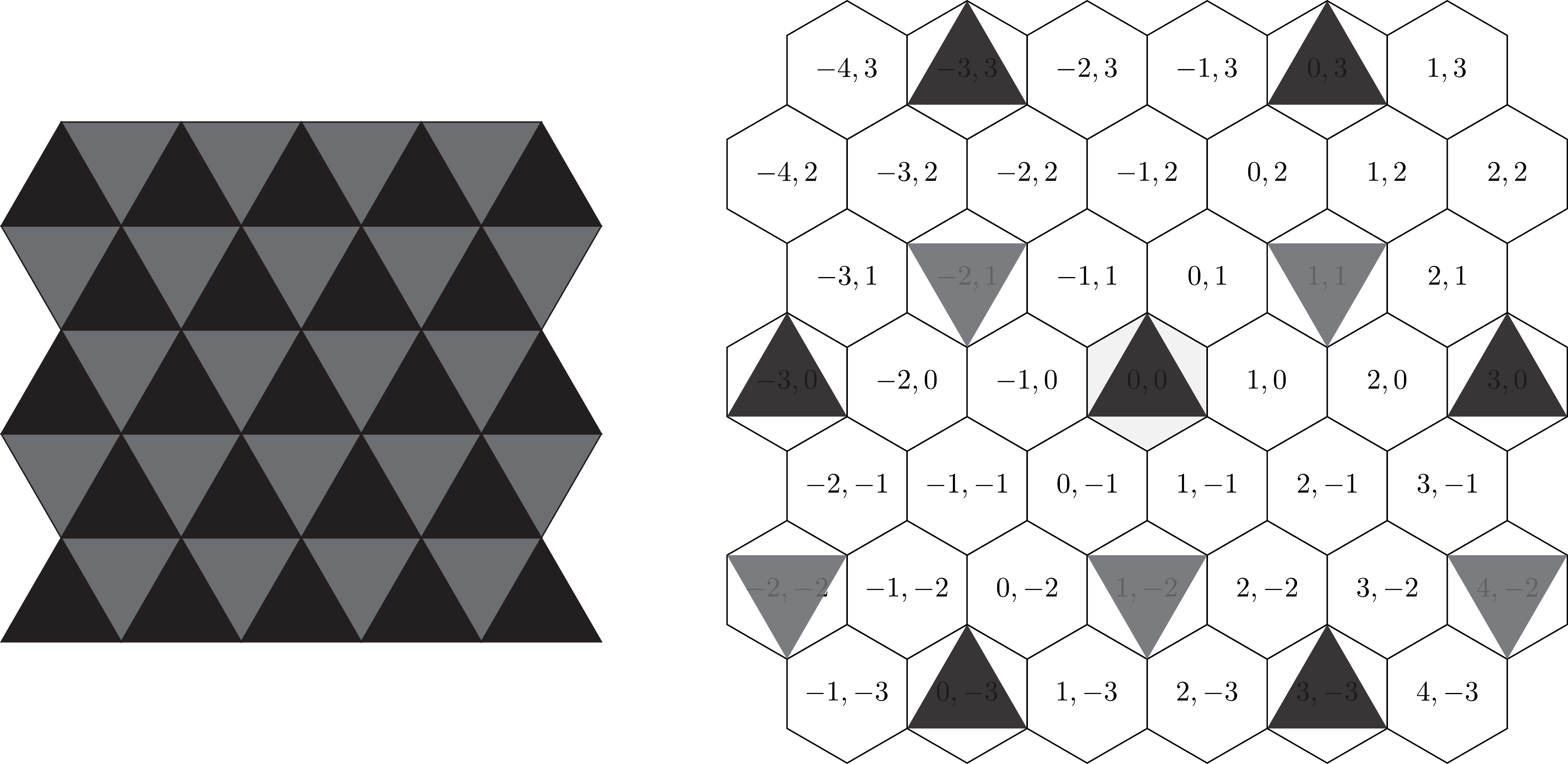}
\end{center}
\caption{\label{fig:trigrid}Polyiamonds can be represented efficiently 
	using a sparse subset of the hexagonal grid.  The conceptual tiling
	on the left is spread out to the coloured cells in the hexagonal
	tiling on the right.}
\end{figure}

\fig{fig:trigrid} shows part of a triangular grid on the left, with 
upward-pointing black triangles and downward-pointing grey triangles.
The illustration on the right shows how triangles are assigned coordinates
in the hexagonal grid.  Every black triangle has coordinates that
are divisible by 3; every grey triangle has coordinates that are
congruent to 1 modulo 3.  Other hexagonal cells are simply left unused.

Like a polyhex, a polyiamond has a maximum of twelve orientations.
Six of these correspond to automorphisms of the black triangle at
$(0,0)$ in \fig{fig:trigrid}, and can be found among the orientation
matrices for hexominoes.  The other six combine one of these six
transformations with a transformation that swaps black and grey
triangles, for example an application of $B$ above followed by a
translation by $(1,-2)$.  Any transformation of a polyiamond can be
represented by a choice of orientation together with a translation
by a vector whose coordinates are divisible by 3.

Neighbourhoods must also be reconsidered in this model.  When computing
haloes we must take into account the \text{12-neighbourhood} of each
cell in a polyiamond $S$, consisting of all cells that share an edge or
a vertex with the given cell.  In \fig{fig:trigrid}, the 12-neighbourhood
of $(0,0)$ consists of every other black and grey triangle shown, together
with one more at $(-2,4)$.  The 12-neighbourhood is fine when computing
haloes and determining adjacency, but not when checking a packing for
holes.  In that case, a flood fill algorithm should move from a given
cell only to the three neighbours with which it shares an edge.

\section{Implementation and results}

I have implemented the data structures and algorithms described here as 
three separate C++ programs, for the three different polyform types.
Each program reads a sequence of polyforms in plain text format,
and produces a text report with the values of $H_c$ and $H_h$ for the
input shapes.  A command line option causes the programs to include, for
each shape, the set of transformations that make up the coronas in the
packings found by the SAT solver.  A separate Python script can read the
shape description and transformations and draw the coronas
that realize the shape's computed Heesch number.

The programs are unable to determine whether a shape tiles the
plane, and must be given known non-tilers as input.  I use software
written by Joseph Myers~\cite{Myers} to enumerate free polyforms
(which are unique up to rotation and reflection) and discard
the shapes that tile.  I then use a separate program to convert from 
the representation Myers uses in his output (a boundary word made up 
of unit steps from an alphabet of evenly spaced directions) to an
area-based representation (coordinates of cells that make up a polyform).

I use the open-source 
CryptoMiniSat library~\cite{CMS} as my SAT solver.  The library
is easy to configure, has a simple C++ API, and performs well in practice.

\begin{table}
\caption{\label{tab:onh}Heesch numbers of $n$-ominoes with no holes 
	in the outer corona}

\begin{center}
\begin{tabular}{crrrrrr}
   $n$ & non-tilers & $H_c=0$ & $H_c=1$ & $H_c=2$ & $H_c=3$ &  \\
   \hline
  7 & 3 & 1 & 2 &  &  &  \\
  8 & 20 & 6 & 14 &  &  &  \\
  9 & 198 & 75 & 122 & 1 &  &  \\
  10 & 1390 & 747 & 642 & 1 &  &  \\
  11 & 9474 & 5807 & 3628 & 39 &  &  \\
  12 & 35488 & 28572 & 6906 & 10 &  &  \\
  13 & 178448 & 149687 & 28694 & 67 &  &  \\
  14 & 696371 & 635951 & 60362 & 58 &  &  \\
  15 & 2721544 & 2598257 & 123262 & 25 &  &  \\
  16 & 10683110 & 10397466 & 285578 & 66 &  &  \\
  17 & 41334494 & 40695200 & 639162 & 130 & 2 &  \\
  18 & 155723774 & 154744331 & 979375 & 68 &  &  \\
  19 & 596182769 & 593856697 & 2325874 & 198 &  &  \\
\end{tabular}
\end{center}
\end{table}

\begin{table}
\caption{\label{tab:oh}Heesch numbers of $n$-ominoes with holes 
	permitted in the outer corona}

\begin{center}
\begin{tabular}{crrrrrr}
   $n$ & non-tilers & $H_h=0$ & $H_h=1$ & $H_h=2$ & $H_h=3$ &  \\
   \hline
  7 & 3 & 0 & 3 &  &  &  \\
  8 & 20 & 0 & 19 & 1 &  &  \\
  9 & 198 & 36 & 157 & 5 &  &  \\
  10 & 1390 & 355 & 1020 & 15 &  &  \\
  11 & 9474 & 2820 & 6544 & 109 & 1 &  \\
  12 & 35488 & 17409 & 18038 & 41 &  &  \\
  13 & 178448 & 100180 & 78048 & 219 & 1 &  \\
  14 & 696371 & 485807 & 210362 & 202 &  &  \\
  15 & 2721544 & 2185656 & 535724 & 164 &  &  \\
  16 & 10683110 & 9300840 & 1381965 & 305 &  &  \\
  17 & 41334494 & 37932265 & 3401701 & 525 & 3 &  \\
  18 & 155723774 & 148955184 & 6768266 & 324 &  &  \\
  19 & 596182769 & 580412188 & 15769814 & 767 &  &  \\
\end{tabular}
\end{center}
\end{table}

\begin{table}
\caption{\label{tab:hnh}Heesch numbers of $n$-hexes with no holes 
	in the outer corona}

\begin{center}
\begin{tabular}{crrrrrr}
   $n$ & non-tilers & $H_c=0$ & $H_c=1$ & $H_c=2$ & $H_c=3$ & $H_c=4$ \\
   \hline
  6 & 4 &  & 3 & 1 &  &  \\
  7 & 37 & 5 & 25 & 6 & 1 &  \\
  8 & 381 & 70 & 264 & 44 & 3 &  \\
  9 & 2717 & 825 & 1822 & 67 & 3 &  \\
  10 & 18760 & 8248 & 10234 & 265 & 13 &  \\
  11 & 116439 & 67644 & 47940 & 817 & 37 & 1 \\
  12 & 565943 & 431882 & 133484 & 567 & 10 &  \\
  13 & 3033697 & 2565727 & 466159 & 1783 & 27 & 1 \\
  14 & 14835067 & 13676416 & 1156793 & 1836 & 22 &  \\
  15 & 72633658 & 69871458 & 2758485 & 3534 & 179 & 2 \\
  16 & 356923880 & 350337478 & 6581529 & 4818 & 54 & 1 \\
  17 & 1746833634 & 1731652467 & 15167876 & 13129 & 161 & 1 \\
\end{tabular}
\end{center}
\end{table}

\begin{table}
\caption{\label{tab:hh}Heesch numbers of $n$-hexes with holes permitted
	in the outer corona}

\begin{center}
\begin{tabular}{crrrrrr}
   $n$ & non-tilers & $H_h=0$ & $H_h=1$ & $H_h=2$ & $H_h=3$ & $H_h=4$ \\
   \hline
  6 & 4 &  & 3 & 1 &  &  \\
  7 & 37 & 4 & 19 & 12 & 2 &  \\
  8 & 381 & 37 & 253 & 84 & 7 &  \\
  9 & 2717 & 434 & 2091 & 185 & 7 &  \\
  10 & 18760 & 4332 & 13766 & 632 & 29 & 1 \\
  11 & 116439 & 38621 & 75783 & 1956 & 73 & 6 \\
  12 & 565943 & 286656 & 277601 & 1652 & 32 & 2 \\
  13 & 3033697 & 1895666 & 1132994 & 4985 & 50 & 2 \\
  14 & 14835067 & 11201813 & 3627594 & 5614 & 46 &  \\
  15 & 72633658 & 61761205 & 10862327 & 9802 & 322 & 2 \\
  16 & 356923880 & 325357916 & 31551809 & 13997 & 156 & 2 \\
  17 & 1746833634 & 1660634503 & 86167750 & 30811 & 569 & 1 \\
\end{tabular}
\end{center}
\end{table}

\begin{table}
\caption{\label{tab:inh}Heesch numbers of $n$-iamonds with no holes 
	in the outer corona}

\begin{center}
\begin{tabular}{crrrrrr}
   $n$ & non-tilers & $H_c=0$ & $H_c=1$ & $H_c=2$ & $H_c=3$ & $H_c=4$ \\
   \hline
  7 & 1 &  & 1 &  &  &  \\
  8 & 0 &  &  &  &  &  \\
  9 & 20 & 11 & 9 &  &  &  \\
  10 & 103 & 44 & 55 & 3 & 1 &  \\
  11 & 594 & 236 & 346 & 11 & 1 &  \\
  12 & 1192 & 826 & 364 & 1 & 1 &  \\
  13 & 6290 & 4360 & 1884 & 24 & 2 &  \\
  14 & 18099 & 14949 & 3141 & 8 &  &  \\
  15 & 54808 & 48108 & 6661 & 39 &  &  \\
  16 & 159048 & 148881 & 10153 & 13 & 1 &  \\
  17 & 502366 & 474738 & 27544 & 83 & 1 &  \\
  18 & 1374593 & 1341460 & 33100 & 33 &  &  \\
  19 & 4076218 & 4001470 & 74689 & 57 & 2 &  \\
  20 & 11378831 & 11282686 & 96091 & 51 & 2 & 1 \\
  21 & 32674779 & 32505745 & 168959 & 73 & 2 &  \\
  22 & 93006494 & 92740453 & 265977 & 62 & 2 &  \\
  23 & 264720498 & 264216706 & 503651 & 140 & 1 &  \\
  24 & 748062099 & 747476118 & 585571 & 384 & 26 &  \\
\end{tabular}
\end{center}
\end{table}

\begin{table}
\caption{\label{tab:ih}Heesch numbers of $n$-iamonds with holes permitted
	in the outer corona}

\begin{center}
\begin{tabular}{crrrrrr}
   $n$ & non-tilers & $H_h=0$ & $H_h=1$ & $H_h=2$ & $H_h=3$ & $H_h=4$ \\
   \hline
  7 & 1 &  & 1 &  &  &  \\
  8 & 0 &  &  &  &  &  \\
  9 & 20 & 7 & 13 &  &  &  \\
  10 & 103 & 33 & 59 & 10 &  & 1 \\
  11 & 594 & 117 & 446 & 30 & 1 &  \\
  12 & 1192 & 495 & 692 & 4 & 1 &  \\
  13 & 6290 & 2639 & 3598 & 51 & 2 &  \\
  14 & 18099 & 10328 & 7745 & 25 &  &  \\
  15 & 54808 & 36965 & 17748 & 91 & 4 &  \\
  16 & 159048 & 124954 & 34058 & 35 & 1 &  \\
  17 & 502366 & 414119 & 88072 & 173 & 2 &  \\
  18 & 1374593 & 1239971 & 134541 & 80 & 1 &  \\
  19 & 4076218 & 3776105 & 299954 & 157 & 2 &  \\
  20 & 11378831 & 10921532 & 457157 & 139 & 2 & 1 \\
  21 & 32674779 & 31831654 & 842947 & 174 & 4 &  \\
  22 & 93006494 & 91551851 & 1454494 & 147 & 2 &  \\
  23 & 264720498 & 262051399 & 2668753 & 343 & 3 &  \\
  24 & 748062099 & 744472222 & 3589353 & 425 & 99 &  \\
\end{tabular}
\end{center}
\end{table}

A SAT solver imposes a small amount of overhead on running time,
because of the need to translate problems from their geometric
origins into Boolean formulas.  However, the benefits of the solver
more than compensate for this added cost.  Human intuition is easily
seduced by the structure of a geometric problem, and that intuition
colours the choice of algorithm used in solving the problem.
Sometimes the resulting algorithms are perfectly fine.  But here,
a ``natural'' approach---walk around the boundary of a shape, gluing
on neighbours, and backtrack when no legal option exists for
continuing---can get stuck in ``backtracking hell''.  An
unavoidable dead end may lurk far out along the boundary of a shape,
with exponentially many (or more!) configurations of neighbours to
be explored along the way, all of which will be rejected.  The
earlier work of Mann and Thomas~\cite{MT2016} attempts to surround
in a fixed order, and they report a number of cases where their
algorithm times out.  A SAT solver has no particular opinion on the
geometric structure of the problem domain.  Its input is an
undifferentiated collection of clauses, and it will take advantage
of any opportunity it can find to narrow the search space, regardless
of order or locality.

I have not attempted to gather full information about the running
times of these programs.  On a single core of a 40-CPU cluster
node with 2.2 GHz Intel Xeon processors,
I can compute the Heesch numbers
of all 1390 non-tiling 10-ominoes in about 220 seconds, on average
about 0.16 seconds per shape.  I have also sampled the running times
on batches of the much larger 17-hexes, and the average per-shape 
computation time is comparable.
Unsurprisingly, the computation time appears to increase exponentially
for shapes with higher Heesch numbers.  For example, shapes with Heesch
number~4 might require 30 seconds to a minute of computation time. 
But because such shapes become progressively more rare as the Heesch
number increases,
the overall effect on computation time is negligible.

Tables~\ref{tab:onh}--\ref{tab:ih} list Heesch numbers for all the 
non-tiling polyforms I
tested, up to 19-ominoes, 17-hexes, and 24-iamonds.  For values of $n$ 
smaller than those shown in the tables, no non-tilers exist.
Permitting holes in the outermost corona offers shapes more freedom to
form coronas.  As a result, the rows of the $H_h$ tables are weighted
slightly more to the right than the corresponding rows of the $H_c$ tables.

\begin{figure}
\begin{center}
\includegraphics[width=5in]{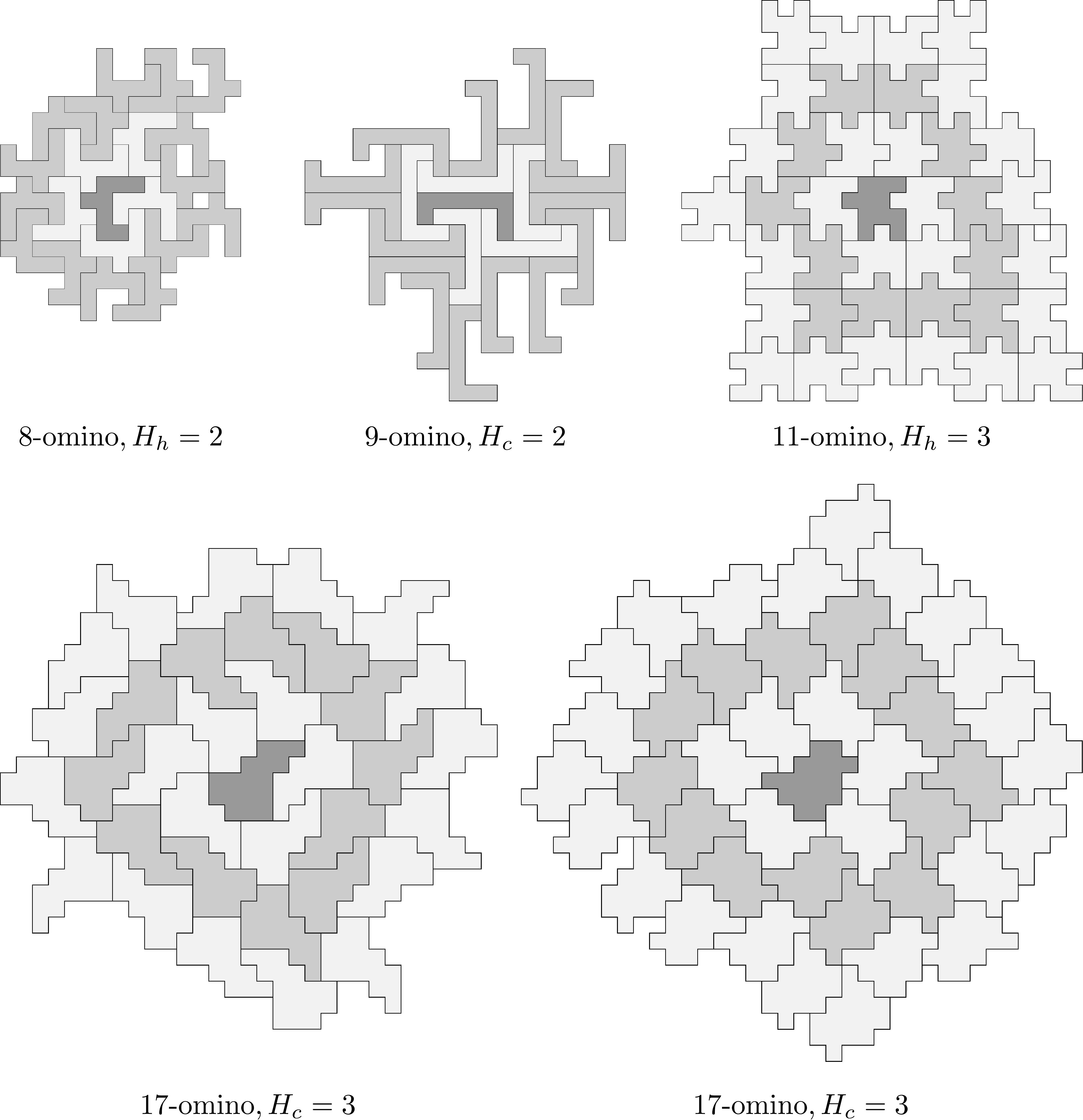}
\end{center}
\caption{\label{fig:ominoes}The smallest polyominoes with Heesch 
	numbers~2 and~3, with and without holes in the outermost corona.
	The 11-omino has a single square hole on the right side of the 
	packing.}
\end{figure}

\begin{figure}
\begin{center}
\includegraphics[width=5in]{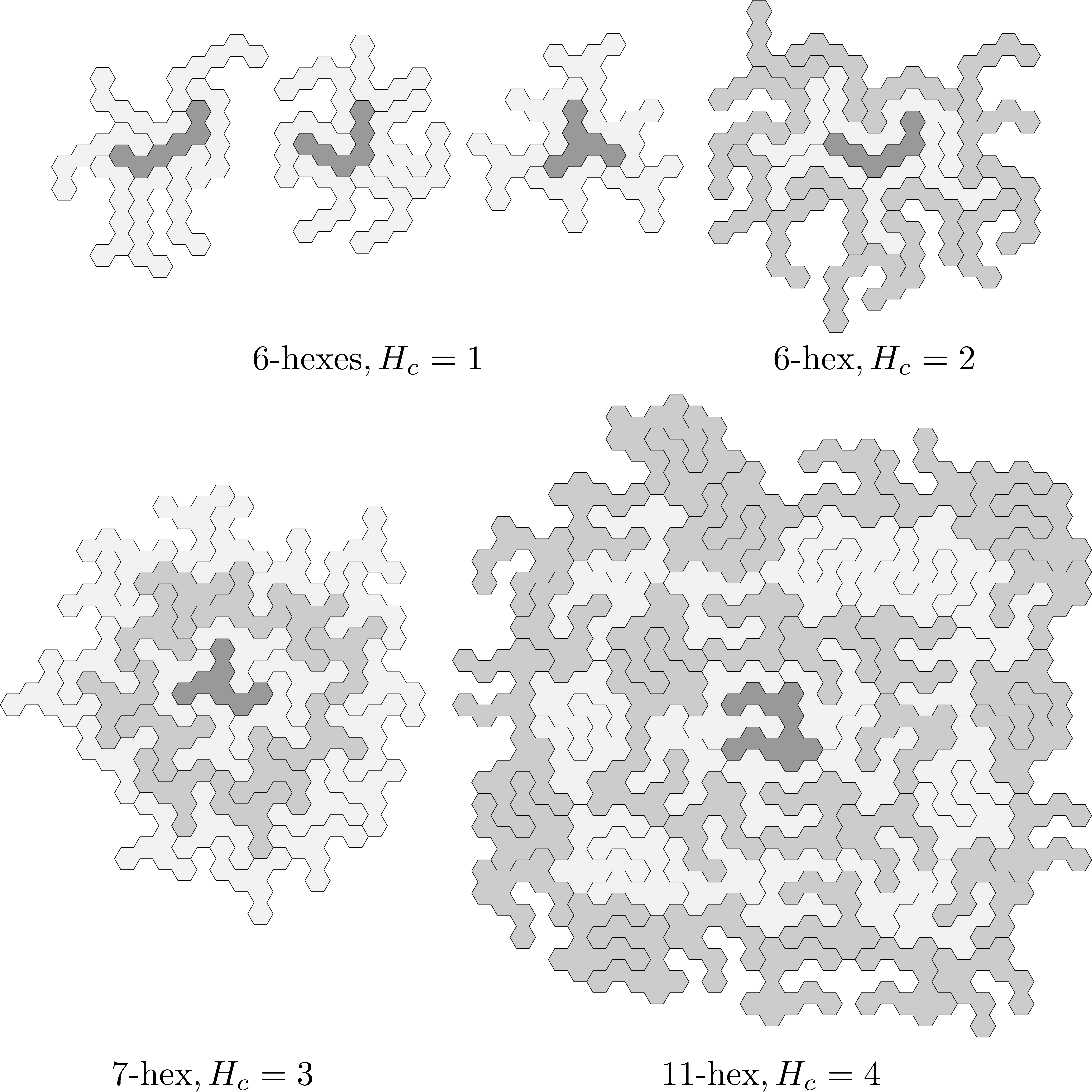}
\end{center}
\caption{\label{fig:hexes}The smallest polyhexes with $H_c=1,2,3,4$.}
\end{figure}

\begin{figure}
\begin{center}
\includegraphics[width=5in]{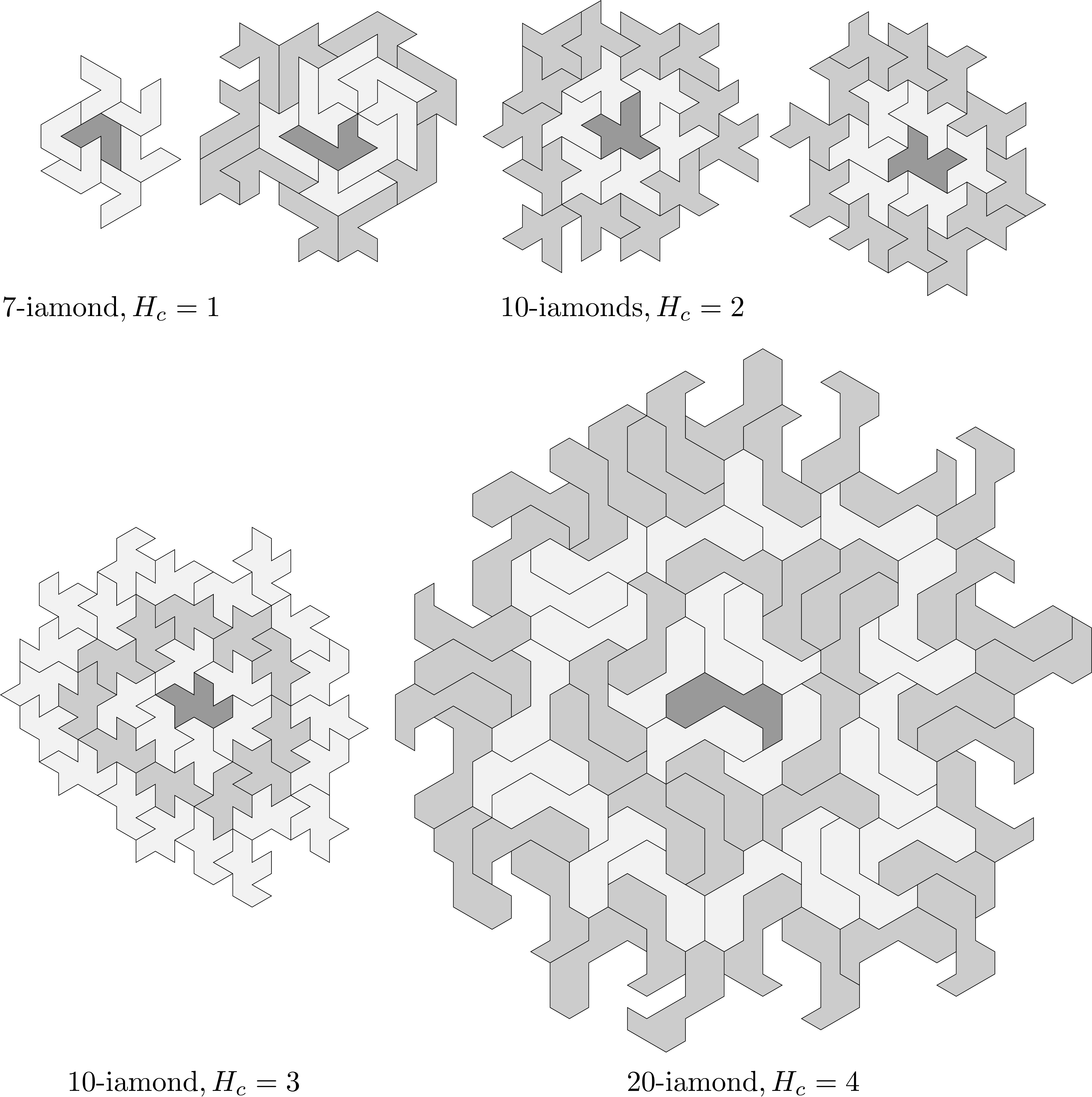}
\end{center}
\caption{\label{fig:iamonds}The smallest polyiamonds with $H_c=1,2,3,4$.}
\end{figure}

Of course, a few highlights deserve to be shared.  I am particularly
interested in the smallest polyforms that exhibit each successive Heesch
number.  \fig{fig:7} already shows the smallest polyominoes with 
$H_c=1$ and $H_h=1$.  \fig{fig:ominoes} shows the smallest polyominoes
with Heesch numbers~2 and~3, both with and without holes. 
In Figures~\ref{fig:hexes} and~\ref{fig:iamonds}, I show the smallest
polyhexes and polyiamonds with hole-free Heesch numbers~1 through~4.  In all
cases, my search did not produce any shapes with Heesch numbers higher
than the ones shown.

\section{Conclusions}

In this article I have demonstrated the effectiveness of recasting the
computation of Heesch numbers within the framework of Boolean satisfiability.
I used a software implementation of this idea to compute Heesch numbers
for a few billion unmarked polyominoes, polyhexes, and polyiamonds.  The
search did not yield any shapes that break previous records for Heesch
numbers, but provides a lot of data that can be used to deepen our
understanding of this intriguing open problem in tiling theory.

The most obvious avenue for future work is to continue the enumeration
to larger polyforms.  However, I am reluctant to do so without
significant performance improvements or insights on narrowing the
set of polyforms to process.  For example, there are more than twice
as many 18-hexes as all the Heesch numbers I have computed so far:
over 8.5 billion of them.  If they require an average of 0.15 seconds
each to process, I estimate that a 120-core cluster would have to
run full-tilt for four months to compute them all.

It would be interesting to reformulate the approach presented here
using binary integer programming~\cite{gurobi} 
instead of Boolean satisfiability.
Some families of clauses might be expressed much more compactly this way.
With satisfiability, if transformed shapes $T_1(S), \ldots, T_m(S)$
all overlap at some cell, then $\binom{m}{2}$ clauses of the form
$s_{T_i,k_i}\rightarrow \neg s_{T_j,k_j}$ are required to rule out
all possible overlaps.  In binary integer programming, the shape
variables would be assigned the integers~0 or~1, and all overlaps
at this cell could be prevented with the single inequality
$s_{T_1,k_1}+\ldots+s_{T_m,k_m}\le 1$.  However, it is unclear whether
this change would boost performance.


Part of the goal of assembling a large corpus of data is to mine it for
patterns.  I do not believe that the tables in this article betray
any obvious patterns in the sizes of polyforms that produce certain 
Heesch numbers.  The general upward trend in each column could be
a simple consequence of the exponential growth in the number of shapes
being classified, and even then the numbers jump around erratically.
But there may be some insight to be gleaned from examinations of the shapes
themselves.  Mann and Thomas refer to ``forced grouping'', in which
tiles in a patch tend to cluster together into larger units~\cite{MT2016}.
I have 
observed this phenomenon in many of my results as well---see for example
the 11-hex patch in \fig{fig:hexes}.  Forced grouping may inspire
strategies for ``amplifying'' the Heesch number of a large shape by
finding a way to decompose it into smaller congruent pieces.

Perhaps the most promising way forward is to consider other families
of shapes.  The techniques in this article could easily be extended
to handle marked polyforms, simply by prohibiting adjacencies that
are not compatible with the markings.  However, it would be crucial to
apply markings to polyforms that tile the plane.  Markings can
only lower an unmarked  shape's Heesch number, making it pointless to add
markings to any of the polyforms presented here. It would therefore
become necessary to check explicitly that a set of markings prevents a 
polyform 
from tiling,
whether based on combinatorial imbalance or a more complex computation.
Of course, it would be interesting to explore the use of a SAT solver
(or integer programming) to check whether a shape tiles the plane.

It may also be possible to extend this work to polyforms that are not
subsets of an ambient grid, like the polyabolos mentioned previously,
or shapes constructed from unions of Penrose rhombs.
In that case we would likely have to do away with haloes and cell variables,
and
use computational geometry to test whether two copies of a shape
are disconnected, adjacent, or overlapping.  The lack of a grid to
organize the plane would incur a heavy cost, but the greater potential
for disorder may pack higher Heesch numbers into smaller shapes.

\section*{Acknowledgments}

Acknowledgments withheld during peer review.


    
{\setlength{\baselineskip}{13pt} 
\raggedright				
\bibliographystyle{plain}
\bibliography{heesch}
}
   
\end{document}